\documentclass[aps,prl,twocolumn,preprintnumbers,10pt]{revtex4-1}
\usepackage{epsf,color,amsmath,amssymb,amsfonts}
\usepackage{enumerate}
\usepackage{hyperref}
\usepackage{graphicx}
\usepackage{psfrag}
\usepackage{dcolumn,bm}

\begin{document}
\onecolumngrid

\preprint{ITP-UU-12/36}

\title{Holographic thermalization with radial flow}

\author{Wilke van der Schee,}

\affiliation{Institute for Theoretical Physics and Institute for Subatomic Physics,
Utrecht University, Leuvenlaan 4, 3584 CE Utrecht, The Netherlands}

\email{w.vanderschee@uu.nl}


\begin{abstract}
\noindent
Recently, a lot of effort has been put into describing  the thermalization
of the quark-gluon plasma using the gauge/gravity duality. In this
context we here present a full numerical solution of the early far-from-equilibrium
formation of the plasma, which is expanding radially in the transverse
plane and is boost invariant along the collision axis. 
This can model the early stage of a head-on relativistic heavy ion collision.
The resulting momentum distribution quickly reaches local equilibrium, after which
they can be evolved using ordinary hydrodynamics. We comment on general implications for
these hydrodynamic simulations, both for central and non-central collisions,
and including fluctuations in the initial state.
\end{abstract}

\maketitle

\noindent \textbf{1. Introduction. }Describing the very early stage
of a relativistic heavy ion collision has remained challenging, in particular when
the state is far-from-equilbrium. In this stage QCD is partially strongly
coupled and impossible to solve in general. 
However, a lot of progress has been made by employing a dual gravitational
description (see \cite{review} and its references), the two seminal examples being the fast thermalization
\cite{Chesler:2008hg,Kovchegov:2010} and the very small viscosity in the consequent hydrodynamic
regime \cite{Policastro:2001}.

These dual descriptions have been studied for near-equilibrium physics
\cite{Policastro:2001,Horowitz:2000,Friess:2008} and also (numerically) far-from-equilibrium \cite{incl,Chesler:2008hg,Beuf:2009cx,Chesler:2010b,Chesler:2011ds,Heller:2011ju,Heller:2012}. Yet, these
studies have always assumed homogeneity in the transverse plane, which
in particular makes it impossible to study radial flow. Recently,
Bantilan, Gubser and Pretorius \cite{Bantilan:2012vu} have performed a simulation having
both longitudinal and radial expansion. However, in their set-up it
was not possible to have boost-invariance in the longitudinal direction
and, more importantly, the evolution was always in the hydrodynamic
regime. In this Letter we will present a numerical simulation having
radial flow in the transverse plane, boost-invariance in the longitudinal
direction, and being far-from-equilibrium initially.

Essentially, we follow the numerical scheme worked out in \cite{Chesler:2010b},
but we assume boost-invariance in the longitudinal direction and allow
for non-trivial radial dynamics in the transverse plane. This keeps
the gravitational problem 2+1 dimensional, which can be solved using
pseudo-spectral evolution. As initial conditions we present two simple
models; the first starts with a blob of energy with a diameter of
approximately 14 fm in vacuum, whereas the second has a blob of about
1 fm in a bath of half the peak energy density. These initial states
can model the overall thermalization of a central collision and the
evolution of an initial fluctuation in such a collision (fluctuations
recently became a topic of much interest, see for instance \cite{Teaney:2011}).
For the bulk metric we started with vacuum AdS, but adapted the near-boundary
coefficients for the energy density and the pressures according to
the Glauber model.

The results of our simulations are both intuitive and encouraging.
Firstly, we find that our geometry thermalizes very quickly, confirming
previous studies. Secondly, our radial velocity profile at the end
of our evolution is similar to typical initial conditions used
for a hydrodynamical evolution. So these radial velocities serve as
a confirmation that current hydrodynamic simulations do not have to
be modified dramatically, but they also provide an improvement for
hydrodynamic evolutions.

\noindent \textbf{2. Holographic model.} As our coordinates in the
field theory it is natural to use proper time $\tau$ and rapidity
$y$, defined by $t=\tau\cosh y$ and $x_{||}=\tau\sinh y$, and angular
coordinates $\rho$ and $\theta$ in the transverse plane. The assumptions
of boost-invariance and rotational symmetry then imply that all functions
are independent of $y$ and $\theta$. In these coordinates the flat
metric of the field theory reads
\begin{equation}
ds_{B}^{2}=-d\tau^{2}+d\rho^{2}+\rho^{2}d\theta^{2}+\tau^{2}dy^{2}.\label{eq:metric-boundary}
\end{equation}
\noindent Given these symmetries and using generalized Eddington-Finkelstein
coordinates, we can write the dual bulk metric as
\begin{eqnarray}
ds^{2}= & -Ad\tau^{2}+\Sigma^{2}(e^{-B-C}dy^{2}+e^{B}d\rho^{2}+e^{C}d\theta^{2})\nonumber \\
 & +2drd\tau+2Fd\rho d\tau,\label{eq:metric}
\end{eqnarray}
\noindent where $A$, $B$, $C$, $\Sigma$ and $F$ are all functions
of $\tau$, $\rho$ and the bulk radial coordinate $r$. Solving Einstein's
equations (with cosmological constant $\Lambda=-6$) order by order
in $r$, demanding that $ds^{2}|_{r=\infty}=r^{2}ds_{B}^{2}$, gives
us the near-boundary expansion of the bulk metric:
\small
\begin{eqnarray}
A & = & r^{2}+\frac{a_{4}(\tau,\rho)}{r^{2}}+O\left(r^{-3}\right),\nonumber\\
B & = & -\frac{2}{3}\log(\tau\rho)+\frac{3r\tau(1-2r\tau)-2}{9r^{3}\tau^{3}}+\frac{b_{4}(\tau,\rho)}{r^{4}}+O\left(r^{-5}\right),\nonumber\\
C & = & -\frac{2}{3}\log(\tau/\rho^{2})+\frac{3r\tau(1-2r\tau)-2}{9r^{3}\tau^{3}}+\frac{c_{4}(\tau,\rho)}{r^{4}}+O(r^{-5}),\nonumber\\
\Sigma & = & \rho^{1/3}\frac{3r\tau(9r\tau(3r\tau(3r\tau+1)-1)+5)-10}{243r^{3}\tau^{11/3}}+O\left(r^{-4}\right),\nonumber\\
F & = & \frac{f_{4}(\tau,\rho)}{r^{2}}+O\left(r^{-3}\right),
\end{eqnarray}
\normalsize
\noindent where in these expressions we fixed a residual gauge freedom
$r\rightarrow r+\xi(\tau,\,\rho)$ by demanding $\partial_{r}A|_{r=\infty}=2r$.
The normalizable modes of the metric, $a_{4}$, $b_{4}$, $c_{4}$
and $f_{4}$, depend on the full bulk geometry and cannot be determined
from a near-boundary expansion. Using the gauge/gravity duality we
can now determine the stress tensor of the dual field theory \cite{deHaro:2000xn}, which
has five independent non-zero components:
\begin{align}
\varepsilon & \equiv-T_{\tau}^{\tau}=-\frac{27}{8\pi^{2}}a_{4},\nonumber \\
s & \equiv T_{\rho}^{\tau}=\frac{9}{2\pi^{2}}f_{4},\nonumber \\
p_{\rho} & \equiv T_{\rho}^{\rho}=\frac{9}{2\pi^{2}}\left(-\frac{1}{6\tau^{4}}-\frac{1}{4}a_{4}+b_{4}\right),\nonumber \\
p_{\theta} & \equiv T_{\theta}^{\theta}=\frac{9}{2\pi^{2}}\left(-\frac{1}{6\tau^{4}}-\frac{1}{4}a_{4}+c_{4}\right),\nonumber \\
p_{y} & \equiv T_{y}^{y}=\varepsilon-p_{\rho}-p_{\theta},
\end{align}
\noindent all functions of $\tau$ and $\rho$, where we put the number
of colors $N_{c}=3$. The conservation of the stress tensor implies
that
\begin{eqnarray}
\partial_{\tau}a_{4} & = & -\frac{12\tau^{4}\left(\rho\left(\tau\partial_{\rho}f_{4}+a_{4}+b_{4}+c_{4}\right)+\tau\, f_{4}\right)-4\rho}{9\rho\tau^{5}},\nonumber \\
\partial_{\tau}f_{4} & = & -\frac{1}{4}\partial_{\rho}a_{4}+\partial_{\rho}b_{4}+\frac{b_{4}-c_{4}}{\rho}-\frac{f_{4}}{\tau}.
\end{eqnarray}

\begin{figure}
\begin{centering}
\includegraphics[width=8cm]{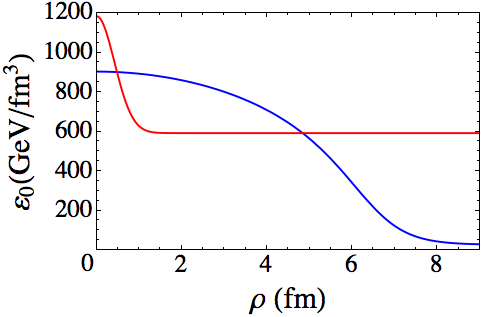}
\par\end{centering}

\caption{The initial energy density profiles at $\tau_{in}=0.12$ fm as a function of the distance
to the origin. The blue curve models a central heavy-ion collision;
the red curve models a fluctuation in such a collision.\label{fig:The-initial-energy}}
\end{figure}

\noindent Our model basically contains two scales: the initial energy
density and the characteristic scale in the radial direction. We can,
however, make use of the scale invariance of the field theory to rescale
our coordinates such that at $\tau=0.6$ fm the energy density at
the origin equals $\varepsilon_{0}=187\,\text{GeV/fm}^{3}$ %
\footnote{One could also choose to match the temperature instead of the energy
density, which is not the same for QCD and our SYM. Here we follow
Gubser \cite{Gubser:2007} and argue that the energy density is more relevant
physically.%
}. We choose this combination to reproduce the final multiplicities
of central heavy-ion collisions at LHC \cite{Niemi:2011}.
For the radial profile we then consider two types of initial conditions,
specified at some small time $\tau_{in}\approx0.12$ fm %
\footnote{In principle, this provides an extra scale, but this initial time
seems small enough not to have a large influence.%
}. The first is a model for a head-on collision, where the shape of
the energy density is provided by the Glauber model, having an approximate
radius of 6.5 fm. The second energy density profile models one
fluctuation in the initial state of such a collision. We take a Gaussian of width
0.5 fm for this profile (see figure \ref{fig:The-initial-energy}). For both initial
conditions we assume that initially there is no radial momentum, such
that $f_{4}(\tau_{in},\,\rho)=0$. 

Importantly, we must also specify the metric functions $B(r,\,\tau_{in},\,\rho)$
and $C(r,\,\tau_{in},\,\rho)$ on a full time-slice of the bulk AdS
geometry. These two functions, together with $a_{4}$, $f_{4}$ and
the Einstein equations, specify the complete metric and its time derivative
on a time-slice \cite{Chesler:2010b}. In principle, these functions should follow from a model describing the very first
weakly coupled stage after the collision, such as the Glauber model
or the Color Glass Condensate. However, these models themselves contain
significant uncertainties and, more importantly, it is not clear how
to map them to this gravitational setting. Therefore we made a simple
choice, where $B$ and $C$ are the same functions as in vacuum AdS,
but with modified $b_{4}$ and $c_{4}$, such that the longitudinal
pressure $p_{y}$ vanishes initially.

Having specified the initial and boundary conditions we can solve
Einstein's equations numerically %
\footnote{The numerical code, results and a movie of the radial velocity can be downloaded at \href{http://www.staff.science.uu.nl/\~schee118/}{www.staff.science.uu.nl/\textasciitilde schee118/}%
}, using essentially the same scheme as in \cite{Chesler:2010b}.
One small difference is the required boundary conditions in the $\rho$ direction, which in this case means
smoothness at the origin and at infinity. As in \cite{Chesler:2010b},
we added a small (3\%) regulator energy density and checked that our
results do not depend on this regulator.

\begin{figure*}
\begin{centering}
\begin{tabular}{cc}
\includegraphics[width=0.42\textwidth]{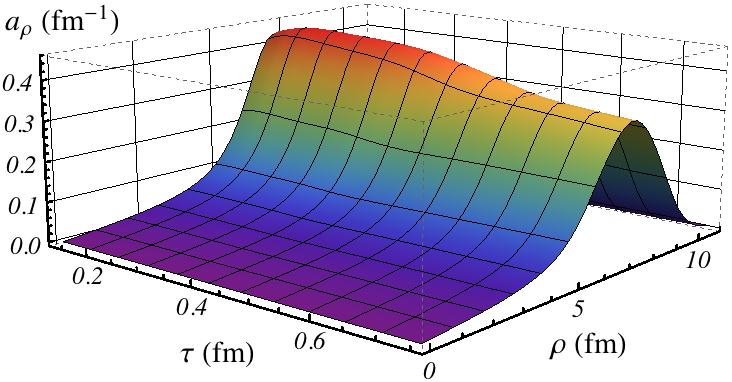}\quad\quad
&  
\quad\quad \includegraphics[width=0.42 \textwidth]{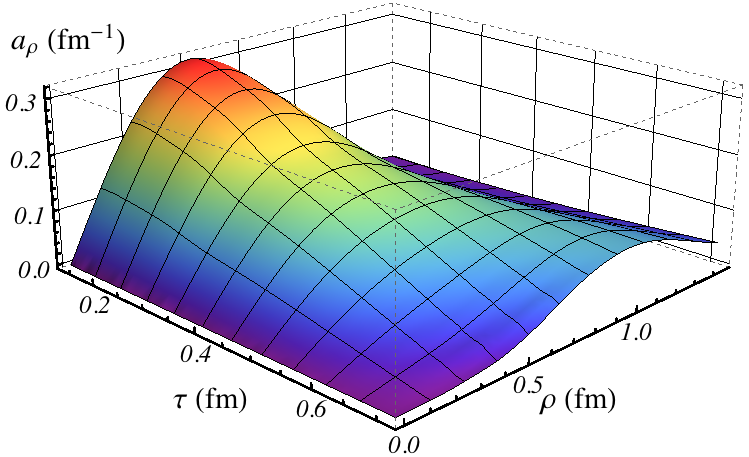}
\\
 (a) & (b)\\ 
\end{tabular}
\end{centering}
\caption{
(a) The radial acceleration of our nucleus model. The acceleration decreases after some time, which is mainly a consequence of the decrease in radial pressure, due to the isotropization. Thereafter the acceleration is quite steady and mainly localized near the boundary of the nucleus.
(b) The radial acceleration of our fluctuation model. Since the bump of energy is much smaller one can clearly see the spreading out and the decrease in acceleration. As will also be clear from figure \ref{fig:velocity-gradient}, this
model reaches a lower radial speed than the model for the nucleus. 
\label{fig:acceleration}}
\end{figure*}

\begin{figure}
\begin{centering}
\includegraphics[width=7.8cm]{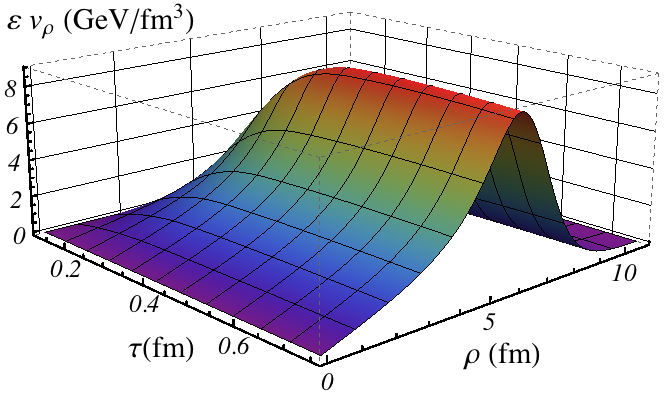}
\par\end{centering}

\caption{The radial velocity times the energy density as a function of proper
time $\tau$ and distance to the origin $\rho$ for our model of a
nucleus. Note that at late times the increasing velocity is almost
exactly compensated by the decreasing energy density (which is due
to the longitudinal expansion). The slope at the origin at the end
of our simulation equals 0.66 GeV/$\text{fm}^{4}$.\label{fig:momentum-flow}}
\end{figure}

\noindent \textbf{3. Results.} After determining the stress tensor
one can extract the radial velocity, defined by the boost after which
there is no momentum flow. Figure \ref{fig:momentum-flow} shows this velocity
times the energy density, which gives a good measure of the momentum
flow. The radial velocity, together with the stress tensor in the
local rest frame, can be used to compute the stress tensor according
to hydrodynamics. Although initially there will not be local equilibrium,
at late times a hydrodynamic expansion is expected to be valid.
It is therefore interesting to compare the actual pressures with the
pressures which follow from a hydrodynamic expansion \cite{Baier:2008,Janik:2005zt}.

\begin{figure}
\begin{centering}
\includegraphics[width=8cm]{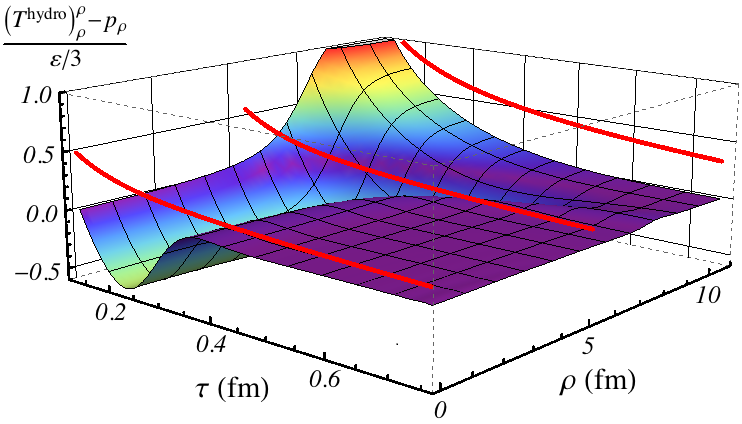}
\par\end{centering}

\caption{The difference between the full non-equilibrium $p_{\rho}$ and the
pressure given by first order hydrodynamics. Although hydrodynamics
applies very quickly, the viscous contribution is still large (shown
by a red lines). The relatively high values for $\rho>7$ fm are a consequence of the very small energy density. For the model of a fluctuation the graph is similar, with equally quick thermalization.\label{fig:hydro}}
\end{figure}

In figure \ref{fig:hydro} we plot the difference of $p_{\rho}$ and
the corresponding first order hydrodynamic prediction of our model
of a nucleus. The stress tensor is excellently described by hydrodynamics 
as soon as $\tau=0.35$ fm. At the border of our nucleus
this is slightly subtler, since the stress tensor is rather small
there, and it becomes comparable to our regulator energy density.
We therefore cannot say too much about this, but the agreement with
hydrodynamics is also there encouraging. We note that in previous
studies \cite{Heller:2011ju,Heller:2012} somewhat larger thermalization times (with respect
to the local temperature) were found, so we expect more exotic initial
conditions in our bulk AdS to give somewhat later thermalization.

In figure \ref{fig:acceleration}b we plot the radial acceleration
of our model of a fluctuation. We notice the acceleration already
decreases considerably during our simulation, in contrast with the
model for the nucleus. Also, the acceleration increases rapidly near
the origin, whereas for the nucleus it is rather narrowly peaked near
the boundary of the nucleus. This means that fluctuations are expected
to spread out rather quickly. Perhaps surprisingly, also
the stress tensor for the fluctuation is governed by hydrodynamics within
0.35 fm.

\noindent \textbf{4. Discussion.} The main motivation for this study
is to provide a description of the far-from-equilibrium stage of heavy-ion
collisions, including non-trivial dynamics in the transverse plane.
While we kept rotational symmetry in the transverse plane, we believe
our study can be used more generally. One reason for this is an old
result in asymptotically flat space \cite{Price:1994pm}, recently studied in asymptotically
AdS \cite{Heller:2012}, that during black hole formation gravity can be well approximated
by linearizing around the final state. We therefore believe that an
initial energy profile with many fluctuations could be well approximated
by superposing the result of our fluctuation presented above.

\begin{figure}
\begin{centering}
\includegraphics[width=7cm]{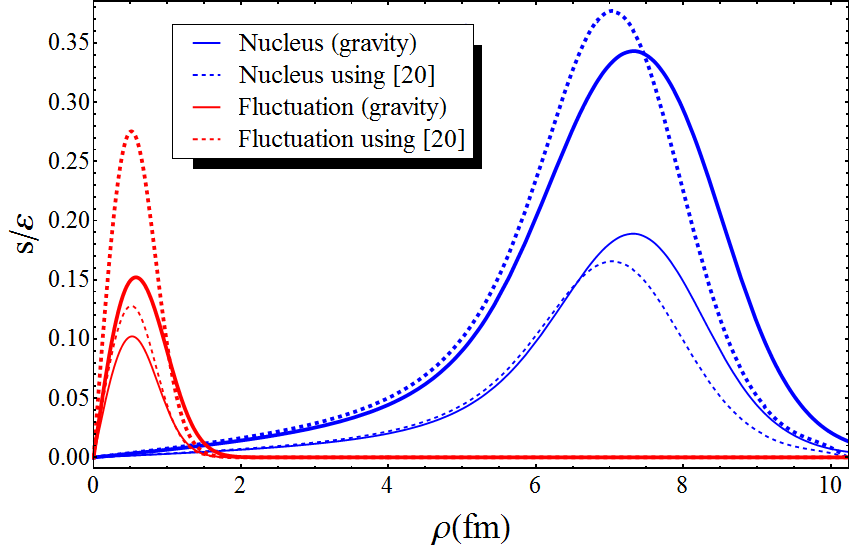}
\par\end{centering}

\caption{Here we plot the momentum flow $s$ divided by the energy density, at time $\tau=0.75$ fm (thick lines) and $\tau=0.4$ fm (thin lines), as a function of $\rho$. The plots compare our gravitational results
with formula \ref{eq:velocitygradient}, which was found in \cite{Pratt:2008}. The two results are remarkably similar, especially at earlier times, and considering the dynamics takes place at very different scales. We suggest formula \ref{eq:velocitygradient} as an initial condition for non-symmetric hydrodynamic simulations, where
a simulation should start at about $\tau=0.4$ fm if fluctuations are present. 
\label{fig:velocity-gradient}}
\end{figure}

Also, it should be possible to use our results for non-central collisions. This can be seen by
comparing with \cite{Pratt:2008}. There, they assume that the anisotropy is independent of $\rho$, the transverse pressures are equal and that the velocity is approximately linear in time.
Without using any hydrodynamics, they used the conservation of the stress tensor to arrive
at the following local formula for the transverse momentum of the stress tensor:
\begin{equation}
\vec{s}/\varepsilon \approx -\frac{\vec{\nabla}_{\perp}\varepsilon_{0}}{2 \varepsilon_{0}}(\tau-\tau_{in}),
\label{eq:velocitygradient}
\end{equation}
\noindent where $\varepsilon_0$ is the initial energy density. This formula (see fig. \ref{fig:velocity-gradient}) works remarkably well at early times and also lateron for the nucleus model. At later times the transverse velocities of fluctuations are smaller, which is due to the decreasing acceleration (displayed in figure \ref{fig:acceleration}b). This result therefore increases confidence in the result of \cite{Pratt:2008}, which can be used in less symmetric situations. When including fluctuations, however, one should hydrodynamics as soons as $\tau=0.4$ fm to get more accurate
results.

Apart from the well-known fact that the gauge/gravity duality
in our setting describes a supersymmetric theory at infinite coupling
and a large number of colors, there is another reason why one should
take care applying our results directly to experimental settings.
Our initial bulk metric should, in principle, follow from the way
the experimental state was created. It is not clear how to do this
and we therefore adopted a simple model, where the metric is basically
close to vacuum. In future work we plan to study the dependence of
the outcomes on this initial state, but preliminary findings suggest
that the results do not depend strongly on the initial bulk metric.

Even though our model is much simpler than a real heavy-ion collision
in QCD, we carried out a far-from-equilibrium calculation at strong
coupling with non-trivial dynamics in the transverse plane. These
results can therefore be very useful as initial conditions
for the hydrodynamic modeling of heavy-ion collisions. Especially
the radial velocity at this initial time was basically unknown and
is hence usually taken to be zero. Our radial velocities would be
a more natural start for a hydrodynamic simulation. Although the final
effect on experimental observables would be rather moderate at the
moment, it will become increasingly important as the experimental
data improve.

\textbf{Acknowledgements.}
It is a pleasure to thank Gleb Arutyunov, Paul Chesler, 
Vivian Jacobs, Ino Karemaker, Thomas Peitzmann, Krishna Rajagopal,
Raimond Snellings and especially Michal Heller, David Mateos and Derek Teaney
for support and useful discussions. We thank the
\emph{Universitat de Barcelona} for hospitality during the completion
of this work. This work is supported under UU grant Foundations of
Science.

\end{document}